\documentstyle[osa,12pt]{revtex}

\input{tcilatex}

\begin{document}
\title{Quantum Switch for Continuous Variable Teleportation}
\author{Junxiang Zhang, Changde Xie, Kunchi Peng}
\address{Institute of Opto-electronics, Shanxi University\\
The Key Laboratory of quantum optics, Ministry of Education, China\\
Shanxi, Taiyuan 030006, P.R.China\\
Tel:+86-351-7010688, Fax:+86-351-7011500\\
Email: kcpeng@sxu.edu.cn}
\maketitle

\begin{abstract}
A novel quantum switch for continuous variables teleportation is proposed.
Two pairs of EPR beams with identical frequency and constant phase relation
are composed on two beamsplitters to produce two pairs of conditional
entangled beams, two of which are sent to two sending stations(Alices) and
others to two receiving stations(bobs). The EPR entanglement initionally
results from two-mode quadrature squeezed state light. Converting the
squeezed component of one of EPR sources between amplitude and phase, the
input quantum state at a sender will be reproduced at two receivers in turn.
The switching system manipulated by squeezed state light might be developed
as a practical quantum switch device for the communication and teleportation
of quantum information.

\end{abstract}

The quantum information technology aims to achieve performance in
communication and computation systems superior to those based on classical
physics by utilizing the nonlocal quantum correlations of entangled states.
It has also significantly improved understanding of the quantum systems
involved on the factual realization of a quantum computer, and raised many
interesting problems such as in the encoding of information\cite{one},
entanglement of states\cite{two}, quantum cryptography\cite{three}, quantum
information manipulation\cite{four}, and quantum communication of
teleportation\cite{five}. One of the most striking features of quantum
information is the quantum teleportations. In the quantum teleportation
scheme, quantum information of an unknown state is faithfully transmitted
from a sender to a remote receiver via an initially shared EPR pairs which
function as a quantum information channel for the faithful transmission.
Quantum teleportation was originally proposed for discrete variable in
finite-dimensional Hilbert space, later it has been successfully developed
from a discrete quantum system into a quantum system for continuous variables%
\cite{six,seven,eight,nine}. Teleportation of optical fields holds great
promise due to the power of the required optical tools and the maturity of
relevant optical communications technology. The quantum teleportation
represents the basic building block of future quantum communication networks
between distant parties\cite{ten}.

To enhance the performance of quantum teleportation system, some attempts
have been made. One of them is to teleport a quantum state from sender to
either of two receivers using the three-particle entanglement at certain
condition of measurement \cite{eleven}. Because of the experimental
difficulties for generating the multiparticle entanglement state\cite{twelve}%
, this teleportation scheme has not been implemented in experiments. In this
paper, we proposed an new novel scheme to teleport an quantum state at Alice
to different two receivers in turn using a two-mode squeezed state as the
quantum switching to manipulate the transmission route. In this scheme, the
EPR entangled beams shared by Alices and Bobs are produced by mixing a pair
of two-mode squeezed state lights with identical frequency and constant
phase relation on two beamsplitters of 50\%. As the usual teleportation
scheme\cite{nine}, one performs a certain joint measurement on the unknown
input quantum state and one of EPR beams from the beamsplitters, and the
Alice's measurement results are transmitted to two BOB's receivers by the
classical lines, then that information is used to transform the second half
of the EPR beams into an output state. Converting one of the two-mode
squeezed state light between amplitude squeezing and phase squeezing, that
can be easily realized by converting the relative phase of the pump field
and injected field of OPA berween 0 and $\pi /2$\cite{thirteen}, the
original unknown input quantum state can be conditionally mimiced at either
of two Bobs in turn, so that the two-mode squeezed state light plays a role
of quantum switching between two spatially separated receivers.

The scheme of the proposed system is shown in Fig.1. A pair of two-mode
squeezed state light are used as the two EPR sources(EPR1 and EPR2). In the
Heisenberg representation, the quadrature amplitude and phase operators $%
\hat{X}$ and $\hat{Y}$ for two modes of each two-mode squeezed state are
expressed as follows\cite{twelve,fourteen}:

\begin{eqnarray}
\hat{X}_{a1} &=&(e^{r_a}\hat{X}_{a1}^{(0)}+e^{-r_a}\hat{X}_{a2}^{(0)})/\sqrt{%
2},\hat{Y}_{a1}=(e^{-r_a}\hat{Y}_{a1}^{(0)}+e^{r_a}\hat{Y}_{a2}^{(0)})/\sqrt{%
2} \\
\hat{X}_{a2} &=&(e^{r_a}\hat{X}_{a1}^{(0)}-e^{-r_a}\hat{X}_{a2}^{(0)})/\sqrt{%
2},\hat{Y}_{a2}=(e^{-r_a}\hat{Y}_{a1}^{(0)}-e^{r_a}\hat{Y}_{a2}^{(0)})/\sqrt{%
2}  \nonumber \\
\hat{X}_{b1} &=&(e^{r_b}\hat{X}_{b1}^{(0)}+e^{-r_b}\hat{X}_{b2}^{(0)})/\sqrt{%
2},\hat{Y}_{b1}=(e^{-r_b}\hat{Y}_{b1}^{(0)}+e^{r_b}\hat{Y}_{b2}^{(0)})/\sqrt{%
2}  \nonumber \\
\hat{X}_{b2} &=&(e^{r_b}\hat{X}_{b1}^{(0)}-e^{-r_b}\hat{X}_{b2}^{(0)})/\sqrt{%
2},\hat{Y}_{b2}=(e^{-r_b}\hat{Y}_{b1}^{(0)}-e^{r_b}\hat{Y}_{b2}^{(0)})/\sqrt{%
2}\text{,}  \nonumber
\end{eqnarray}
where the subscripts $a_1$, $a_2$ and $b_1$, $b_2$ are designated for the
two modes in the two-mode squeezed state from EPR1 and EPR2 respectively.
The subscript ``(0)'' denotes initial coherent modes, $r_a$ and $r_b$ are
the correlation parameters between $a_1$ and $a_2$ as well as $b_1$ and $b_2$
mode. Under the limit of perfect correlation $r_a\rightarrow \infty $ and $%
r_b\rightarrow \infty $, the two two-mode squeezed states approach the
eigenstates of $\hat{X}_{a(b)1}-\hat{X}_{a(b)2}$ and $\hat{Y}_{a(b)1}+\hat{Y}%
_{a(b)2}$, which corresponding to two perfect EPR pairs having the
quadrature amplitude correlation and quadrature phase anticorrelation.
Otherwise, if $r_a\rightarrow -\infty $ and $r_b\rightarrow -\infty $, it
approach the eigenstates of $\hat{X}_{a(b)1}+\hat{X}_{a(b)2}$ and $\hat{Y}%
_{a(b)1}-\hat{Y}_{a(b)2}$, which are also the perfect EPR state with the
anticorrelated quadrature amplitude and correlated quadratures phase.

In order to perform a teleportation switching, the sender has to share
simultaneously the entanglement with two receivers. Initially, mode $a_1$
shares the entanglement with the mode $a_2$ and mode $b_1$ shares the
entanglement with the mode $b_2$, then we mix the mode $a_1$ and $a_2$ with
the mode $b_1$ and $b_2$ at the beamsplitters BS1 and BS2 respectively. The
output modes of the two beamsplitters are:

\begin{eqnarray}
\hat{X}_3 &=&(\hat{X}_{a1}-\hat{X}_{b1})/\sqrt{2},\hat{Y}_3=(\hat{Y}_{a1}-%
\hat{Y}_{b1})/\sqrt{2} \\
\hat{X}_4 &=&(\hat{X}_{a1}+\hat{X}_{b1})/\sqrt{2},\hat{Y}_4=(\hat{Y}_{a1}+%
\hat{Y}_{b1})/\sqrt{2}  \nonumber \\
\hat{X}_5 &=&(\hat{X}_{a2}+\hat{X}_{b2})/\sqrt{2},\hat{Y}_5=(\hat{Y}_{a2}+%
\hat{Y}_{b2})/\sqrt{2}  \nonumber \\
\hat{X}_6 &=&(\hat{X}_{a2}-\hat{X}_{b2})/\sqrt{2},\hat{Y}_4=(\hat{Y}_{a2}-%
\hat{Y}_{b2})/\sqrt{2}\text{.}  \nonumber
\end{eqnarray}

Therefore we have:

\begin{eqnarray}
\hat{X}_3-\hat{X}_5 &=&[(\hat{X}_{a1}-\hat{X}_{a2})-(\hat{X}_{b1}+\hat{X}%
_{b2})]/\sqrt{2} \\
\hat{Y}_3+\hat{Y}_5 &=&[(\hat{Y}_{a1}+\hat{Y}_{a2})-(\hat{Y}_{b1}-\hat{Y}%
_{b2})]/\sqrt{2}  \nonumber \\
\hat{X}_3-\hat{X}_6 &=&[(\hat{X}_{a1}-\hat{X}_{a2})-(\hat{X}_{b1}-\hat{X}%
_{b2})]/\sqrt{2}  \nonumber \\
\hat{Y}_3+\hat{Y}_6 &=&[(\hat{Y}_{a1}+\hat{Y}_{a2})-(\hat{Y}_{b1}+\hat{Y}%
_{b2})]/\sqrt{2}\text{.}  \nonumber
\end{eqnarray}

It is obviously that the mode $a_3$ will be entangled with the mode $a_5$ if
one of the two-mode squeezed states is the eigenstate of $\hat{X}_{a1}-\hat{X%
}_{a2}$ and $\hat{Y}_{a1}+\hat{Y}_{a2}$(i.e. $r_a\rightarrow \infty $), and
the other one is the eigenstate of $\hat{X}_{b1}+\hat{X}_{b2}$ and $\hat{Y}%
_{b1}-\hat{Y}_{b2}$(i.e. $r_b\rightarrow -\infty $). The mode $a_3$ will be
entangled with the mode $a_6$ at the conditions of $r_a\rightarrow \infty $, 
$r_b\rightarrow \infty $ or $r_a\rightarrow -\infty $, $r_b\rightarrow
-\infty $. According to the different entanglement condition the quantum
information from a sender can be controllably transmitted to Bob1 or Bob2.

In Alice's station an unknown input quantum state represented by the
quadrature operators $\hat{X}_{in}$ and $\hat{Y}_{in}$ is superposed with $%
a_3$ mode at the beamsplitter BS3. Two balanced homodyne detectors $D_{Xc}$
and $D_{Yc}$ are used to measure the observables of the quadrature amplitude 
$\hat{X}_c=(\hat{X}_{in}-\hat{X}_3)/\sqrt{2}$ of one output of BS3 and the
quadrature phase $\hat{Y}_c=(\hat{Y}_{in}+\hat{Y}_3)/\sqrt{2}$ of the other
output of BS3 . The resulting classical outcomes are scaled by the operators:

\begin{eqnarray}
\hat{X}_c &=&[\hat{X}_{in}-(\hat{X}_{a1}-\hat{X}_{b1})/\sqrt{2}]/\sqrt{2} \\
\hat{Y}_c &=&[\hat{Y}_{in}+(\hat{Y}_{a1}-\hat{Y}_{b1})/\sqrt{2}]/\sqrt{2}%
\text{.}  \nonumber
\end{eqnarray}

Due to the entanglement between the modes $a_3$ and $a_5$, or $a_3$ and $a_6$
at different conditions, the measurements lead to that the mode $a_5$ or $%
a_6 $ collapses into a state which differs from the unknown input state in
the classical phase-space displacement. Thus for the possibility to recover
the input state at the two locations, each of the classical outcomes is
divided into two same parts with RF power splitters(RF1 and RF2)\cite
{fifteen}, and the classical information is sent separately to the two
remote locations for performing appropriate displacements on modes $\hat{a}%
_5 $ and $\hat{a}_6$:

\begin{eqnarray}
\hat{a}_5 &\rightarrow &\hat{a}_5^{out}=\hat{a}_5+\sqrt{2}g_1\hat{a}_c \\
\hat{a}_6 &\rightarrow &\hat{a}_6^{out}=\hat{a}_6+\sqrt{2}g_2\hat{a}_c\text{,%
}  \nonumber
\end{eqnarray}
where $\hat{a}_c$ represents the annihilation operator of coherent state
generating from the classical results $\hat{a}_c=\hat{X}_c+i\hat{Y}_c$.

According to Eqs.(1) (2) (4) and (5), the outgoings of two modes become:

\begin{eqnarray}
\hat{a}_5^{out} &=&g_1\hat{a}_{in}+\frac{1+g}2[e^{-r_a}(-\hat{X}_{a2}^{(0)}+i%
\hat{Y}_{a1}^{(0)})+e^{r_b}(\hat{X}_{b1}^{(0)}-i\hat{Y}_{b2}^{(0)})] \\
&&+\frac{1-g}2[e^{r_a}(\hat{X}_{a1}^{(0)}-i\hat{Y}_{a2}^{(0)})+e^{-r_b}(-%
\hat{X}_{b2}^{(0)}+i\hat{Y}_{b1}^{(0)})]  \nonumber \\
\hat{a}_6^{out} &=&g_2\hat{a}_{in}+\frac{1+g}2[e^{-r_a}(-\hat{X}_{a2}^{(0)}+i%
\hat{Y}_{a1}^{(0)})+e^{-r_b}(\hat{X}_{b2}^{(0)}-i\hat{Y}_{b1}^{(0)})] 
\nonumber \\
&&\frac{1-g}2[e^{r_a}(\hat{X}_{a1}^{(0)}-i\hat{Y}_{a2}^{(0)})+e^{r_b}(-\hat{X%
}_{b1}^{(0)}+i\hat{Y}_{b2}^{(0)})]\text{,}  \nonumber
\end{eqnarray}
where the parameters $g_1$ and $g_2$ describe the normalized gain of two
teleportation processes from a sender to Bob1 and Bob2.

The Eq(6) show that both the output modes $\hat{a}_5^{out}$ and $\hat{a}%
_6^{out}$ contain some information about the teleported state but it is not
exactly the input state due to some additional noise from the quantum
channels.

For the case of ideal measurement process $g_1=1$ and $r_a\rightarrow \infty 
$, $r_b\rightarrow -\infty $ (or $r_a\rightarrow -\infty $, $r_b\rightarrow
\infty $). The Eq(6) becomes $\hat{a}_5^{out}=\hat{a}_{in}$. So the perfect
quantum teleportation is accomplished at the receiver Bob1.

If $g_2=1$ and $r_a\rightarrow \infty $, $r_b\rightarrow \infty $ (or $%
r_a\rightarrow -\infty $, $r_b\rightarrow -\infty $), we have $\hat{a}%
_6^{out}=\hat{a}_{in}$, then the unknown quantum state is perfectly mimiced
at the receiver Bob2.

The fidelity quantifying the quality of teleportation is defined for a
coherent input state by\cite{sixteen,seventeen}:

\begin{eqnarray}
F &=&\frac 2{\sqrt{(\left\langle \delta ^2\hat{X}_{out}\right\rangle
+1)(\left\langle \delta ^2\hat{Y}_{out}\right\rangle +1)}}  \nonumber \\
&&\exp [-2\frac{(1-g)^2\left| \alpha _{in}\right| ^2}{\sqrt{(\left\langle
\delta ^2\hat{X}_{out}\right\rangle +1)(\left\langle \delta ^2\hat{Y}%
_{out}\right\rangle +1)}}]\text{,}
\end{eqnarray}
where $\delta ^2\hat{X}_{out}$ and $\delta ^2\hat{Y}_{out}$ are the variance
of quadrature amplitude and phase of the output mode. Using Eqs(6), they are
given by:

\begin{eqnarray}
\left\langle \delta ^2\hat{X}_{out}\right\rangle &=&g_{1(2)}^2\langle \delta
^2\hat{X}_{in}\rangle +(\frac{1+g}2)^2[e^{-2r_a}+e^{\pm 2r_b}]+(\frac{1-g}2%
)^2[e^{2r_a}+e^{\mp 2r_b}]  \nonumber \\
\left\langle \delta ^2\hat{Y}_{out}\right\rangle &=&g_{1(2)}^2\left\langle
\delta ^2\hat{Y}_{in}\right\rangle +(\frac{1+g}2)^2[e^{-2r_a}+e^{\pm 2r_b}]+(%
\frac{1-g}2)^2[e^{2r_a}+e^{\mp 2r_b}]\text{,}
\end{eqnarray}
in Eq.(8), the symbol ``$\pm $'' and ``1(2)'' represent that the
teleportation is accomplished at the output mode of $\hat{a}_5$ or $\hat{a}%
_6 $ respectively.

In the classical system without quantum correlation $r_a=0$ and $r_b=0$, we
obtain $F=1/2$ for the normalized gain $g_{1(2)}=1$, so the classical limit
of teleportation in this system keeps the same with the usual teleportation
system for continuous variables\cite{seventeen}. According to eq.(7), to
meet the requirement of the quantum teleportation $F>1/2$, only requires
that either of the two the initial light fields from EPR1 and EPR2 is a
two-mode squeezed state, even while the other one is a coherent state light.
But if one wants to get high fidelity two of them should have high squeezing.

For experiments the most important work is to establish two EPR beam sources
with identical frequency and constant phase relation. Two same degenerate%
\cite{nine} or nondegenerate optical parametric amplifiers\cite
{eighteen,nineteen}(DOPA or NOPA) pumped by a same laser can be used to
produce the needed two two-mode squeezed states. The mature parametric
technique is beneficial to complete the proposed prototype. The correlation
relation between two modes of EPR pairs can be manipulated by converting the
relative phase between pump field and injected signal field of OPA between 0
and $\pi $. For the parametric deamplification (the pump field and the
injected field are in phase of 0), the two-mode amplitude squeezing is
completed which corresponding to the EPR beams with the quadrature amplitude
correlation and quadrature phase anticorrelation between two modes\cite
{twenty}. And for a polarization nondegenerate parametric amplification( the
pump field and the injected field are out of phase i.e. the relative phase
is $\pi $), the two-mode phase squeezing is obtained which corresponding to
the quadrature amplitude anticorrelation and quadrature phase correlation
EPR beams\cite{eighteen,nineteen,twenty-one,twenty-two}.

In conclusion, we propose an quantum switching system for sending
controllably an unknown quantum state to either of two remote receivers. The
control condition is only to convert the squeezed component of one of two
two-mode squeezed states between its quadrature amplitude and phase. The
conditional teleportation system might be developed as a practical quantum
switching in future quantum communication. The well-known optical parametric
technique provides great convenience for its experimental demonstration.

{\bf ACKNOWLEDGMENTS}

This research is supported by the National Natural Science Foundation of
China (No.69978013) and the Oversea Youth Scholar Collabration
Foundation(No.69928504).


\begin{references}
\bibitem{one}  J.Preskill, e-print quant-ph/9904022.

\bibitem{two}  A.Zeilinger, Phys.World 11, 35(1998).

\bibitem{three}  W.Tittel, G.Ribordy and N.Gisin, Phys.World 11, 41(1998).

\bibitem{four}  M.A.Nielsen, Phys.Re.Lett., 83, 436(1999).

\bibitem{five}  C.H.Bennett, G.Brassard, C.Crepeau, Phys.Rev.Lett., 70,
1895(1993).

\bibitem{six}  D.Bouwmeester, J.W.Pan, K.Mattle, M.Edibl, H.Weinfurter and
A.Zeilinger, Nature, 390, 575(1997).

\bibitem{seven}  D.Boschi, S.Branca, F.De Martini, L.Hardy and S.Popescu,
Phys.Rev.Lett., 80, 1121(1998).

\bibitem{eight}  L.Vaidman, Phys.Rev.A49, 1473(1994); S.L.Braunstein,
H.J.Kimble, Phys.Rev.Lett., 80, 869(1998).

\bibitem{nine}  A.Furusawa, J.L.Sorensen, S.L.Braunstein, C.A.Fuchs,
H.J.Kimble, E.S.Polzik, Science, 282, 706(1998).

\bibitem{ten}  S.Bose, V.Vedral and P.L.Knight, Phys.Rev.A57, 822(1998).

\bibitem{eleven}  A.Karlsson and M.Bourennane, Phys.Rev.A, 58, 4394(1998).

\bibitem{twelve}  P.van Loock and S.L.Braunstein, Phys.Rev.Lett., 84,
3482(2000).

\bibitem{thirteen}  S.F.Pereira, Z.Y.Ou and H.J.Kimble, Phys.Rev.A, 62,
042311(2000).

\bibitem{fourteen}  D.F.Walls and G.J.Milburn, Quantum Optics(Springer
Verlag), Berlin, 1994.

\bibitem{fifteen}  J.Zhang, K.C.Peng, will be printed in PRA in November.

\bibitem{sixteen}  S.L.Braunstein, C.A.Fuchs, and H.J.Kimble, J.Mod.Opt.,
47,267(2000); S.L.Braunstein, C.A.Fuchs, and H.J.Kimble, e-print,
quant-ph/9910030.

\bibitem{seventeen}  R.E.S.Polkinghorne and T.C.Ralph, Phys.Rev.Lett., 83,
2095(1999).

\bibitem{eighteen}  Z.Y.Ou, S.F.Pereira, H.J.Kimble, and K.C.Peng,
Phys.Rev.Lett., 68, 3663(1992).

\bibitem{nineteen}  Y.Zhang, H.Wang, X.Y.Li, J.T.Jing, C.D.Xie, K.C.Peng,
Phys.Rev.A, 62, 023813(2000).

\bibitem{twenty}  Y.Zhang, H.Su, C.D.Xie, K.C.Peng, Phys.Lett.A, 259,
171(1999).

\bibitem{twenty-one}  K.Schneider, R.Bruckmeier, H.Hansen, S.Schiller and
J.Mlynek, Opt.Lett., 21, 1396(1996).

\bibitem{twenty-two}  Y.Aharonov et al., Ann.Phys.(N.Y.), 39, 498(1966).
\end{references}
\end{document}